\UseRawInputEncoding
\documentclass[superscriptaddress, notitlepage, reprint,twocolumn]{revtex4-1}
\usepackage[english]{babel}
\usepackage{amsmath,amsthm,amssymb,lipsum}
\usepackage{amsfonts}
\usepackage[pdfborder={0 0 0}, colorlinks=true, urlcolor=blue, linkcolor=blue, citecolor=blue]{hyperref}
\usepackage{color}
\usepackage{graphicx}
\usepackage{float}
\usepackage{subfigure}
\usepackage{lipsum}
\usepackage{epstopdf}
\usepackage{dcolumn}
\usepackage{mathrsfs}
\begin{document}

\title{Nonreciprocal Dispersive Coupling for Quantum Sensing}
\author{Dong  Xie}
\email{xiedong@mail.ustc.edu.cn}
\author{Chunling Xu}

\affiliation{College of Science, Guilin University of Aerospace Technology, Guilin, Guangxi 541004, People's Republic of China}

\begin{abstract}
Dispersive coupling is widely utilized for quantum information readout. Most prior studies have concentrated on reciprocal dispersive coupling. Here, we further construct nonreciprocal dispersive coupling and apply it to quantum sensing. For cavity photon number measurement, nonreciprocal dispersive coupling delivers higher precision than its reciprocal counterpart, and this advantage grows more pronounced with an increase in photon number. When directly measuring the single-photon driving strength, however, nonreciprocal dispersive coupling shows no superiority over reciprocal dispersive coupling. By converting the information of driving strength into cavity photon numbers via our proposed strategy, nonreciprocal dispersive coupling again outperforms reciprocal dispersive coupling in precision, with the advantage becoming increasingly significant at larger driving strength. This work presents a novel method to boost quantum sensing and enable the fabrication of ultra-precise quantum sensors.
\end{abstract}
\maketitle

\section{Introduction}
Dispersive coupling occurs in the regime where the qubit is largely detuned from the bosonic mode, such as, cavity mode. Dispersive coupling has been widely applied in the field of quantum information, such as readout of Majorana qubits~\cite{lab1}, adiabatic phases~\cite{lab2}, quantum circuits~\cite{lab3}, and temperature~\cite{lab4}.

 If an interaction exhibits directionality, it is defined as a nonreciprocal interaction. In such cases, one system can alter the dynamics of another through nonreciprocal coupling, with no reciprocal effect in return. In the classical regime, effective non-Hermitian Hamiltonians can be adopted to describe directional interactions. However, in the quantum regime, a description that preserves probability conservation and accounts for quantum fluctuations is required. Quantum nonreciprocity can exist without classical nonreciprocity~\cite{lab5,lab6}. The theory of cascaded quantum systems~\cite{lab7,lab8} offers a standard framework for quantum nonreciprocal interactions, which requires an effective broken time-reversal symmetry.
A cascaded network of resonant qubits~\cite{lab9} has been extensively explored. The nonreciprocal coupling between two qubits is achieved by coupling to a directional waveguide~\cite{lab10}. Such resonant nonreciprocal coupling can be realized in chiral quantum
optical platforms~\cite{lab11} and waveguide circuit quantum electrodynamics (QED)~\cite{lab12,lab13,lab14}.

Further research~\cite{lab15} demonstrates that nonreciprocity can also be obtained without breaking time-reversal symmetry. A basic gauge symmetry present in
any Lindblad master equation can be utilized to obtain nonreciprocity. A typical representative of this approach stems from dispersive coupling, which can be derived via Lindblad dissipation operator, $D[ae^{i\theta\sigma_z}]$, which enables the cavity mode $a$ to act on the qubit $\sigma_z$, with no reciprocal influence in the reverse direction.
Recently, a nonreciprocal dispersive coupling between a transmon qubit and a superconducting cavity has been realized in experiment~\cite{lab16}.

The nonreciprocal couplings in the cascaded quantum systems has been applied in many fields~\cite{lab17,lab18,lab19,lab20,lab21}.  In the field of quantum sensing,  Ref.~\cite{lab22} firstly showed that nonreciprocity can enhance dispersive measurement. In our prior work~\cite{lab23}, we systematically compared how perfect nonreciprocal coupling outperforms reciprocal coupling in improving the precision of quantum measurements. Therefore, an important question arises: does nonreciprocal dispersive coupling retain its quantum-sensing superiority analogous to nonreciprocal coupling in cascaded systems?

In this article, we provide a definite answer and make a detailed comparison of the magnitude of such advantages. We first construct the simplest nonreciprocal dispersive coupling system via adiabatic elimination. When measuring the cavity photon number, we find that nonreciprocal dispersive coupling yields higher measurement precision than its reciprocal counterpart. Moreover, this superiority increases exponentially with the cavity photon number. When directly measuring the single-photon driving strength, nonreciprocal dispersive coupling shows no superior performance compared with reciprocal dispersive coupling. We propose a novel measurement strategy that fully exploits the merits of nonreciprocal dispersive coupling, thus achieving higher measurement precision than that of its reciprocal counterpart.

The rest of this article is arranged as follows. In section II, nonreciprocal dispersive coupling between cavity mode and qubit is realized via adiabatic elimination of the intermediate bosonic mode. In section III, we derive the measurement uncertainty of cavity photon number for reciprocal and nonreciprocal dispersive coupling systems by employing Pauli operators. The optimal measurement precision is obtained by quantum Fisher information in section IV. In section V, we make a detailed comparison of the advantages of nonreciprocal dispersive coupling over its reciprocal counterpart in cavity photon number measurement precision. In section VI, we conduct a detailed analysis on the advantages of nonreciprocal dispersive coupling in enhancing the measurement  precision of single-photon driving strength. We make a conclusion and outlook in section VII. Detailed derivations are presented in Appendix A-D.

\section{Nonreciprocal Dispersive Coupling }

\begin{figure}[h]
\includegraphics[scale=0.27]{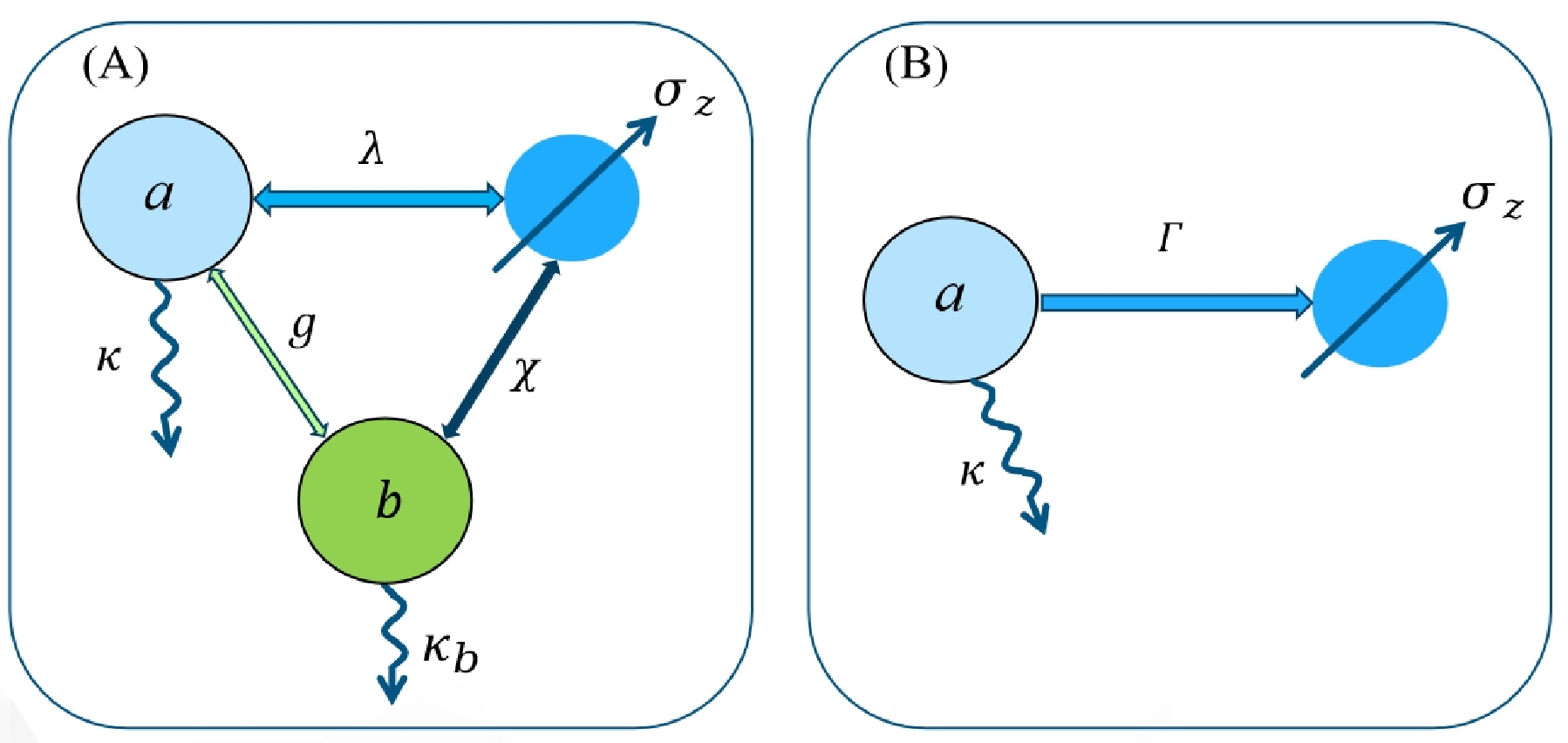}
 \caption{\label{fig.1} Schematic diagram of nonreciprocal dispersive coupling. (A) The cavity mode $a$ is coupled to the bosonic mode $b$ with the coupling strength denoted by $g$. The qubit couples to the cavity mode $a$ and bosonic $b$, with dispersive coupling strengths $\lambda$ and $\chi$, respectively. The dissipation rate suffered by the bosonic mode $b$ is much higher than that of the cavity mode $a$, $\kappa_b\gg(\kappa, g)$. (B) Schematic diagram of the equivalent nonreciprocal coupling between the cavity and the qubit. After fully eliminating bosonic mode $b$ adiabatically, the cavity mode $a$ couples to the qubit $\sigma_z$ via  nonreciprocal dispersive coupling with strength $\Gamma$.}
\end{figure}

We consider a system where a qubit interacts with a cavity via dispersive coupling, as shown in Fig.~\ref{fig.1}. The dispersive Hamiltonian is described by ($\hbar=1$)~\cite{lab24}
\begin{align}
{H_s}=\omega {a}^\dagger{a}+\frac{1}{2}\omega_q{\sigma}_z+\lambda{\sigma}_z{a}^\dagger{a},\label{eq:1}\tag{1}
\end{align}
where $a$ $(a^\dagger)$ represent annihilation (creation) operators of the cavity mode with the frequency $\omega$, $\sigma_z$ denotes the Pauli matrix of the qubit with the transition frequency $\omega_q$, and $\lambda$ denotes the dispersive coherent coupling strength between the qubit and the cavity mode.
The nonreciprocal dispersive coupling can be obtained by introducing an additional bosonic mode. The total Hamiltonian is described by
\begin{align}
H_T=H_s+\omega_b {b}^\dagger{b}+g(a^\dagger b+b^\dagger a)+\chi\sigma_zb^\dagger b,\label{eq:2}\tag{2}
\end{align}
where $b$ $(b^\dagger)$ represent annihilation (creation) operators of the bosonic mode with frequency $\omega_b$, $g$ denotes the coherent coupling strength between the cavity mode $a$ and bosonic mode $b$, and $\chi$ stands for the dispersive coupling strength between the qubit and bosonic mode $b$.
 When the dissipation rate of bosonic mode $b$ far exceeds that of cavity mode $a$, the bosonic mode $b$ can be adiabatically eliminated. In a rotating reference frame, the evolution dynamics of the cavity mode and qubit is described by the Markovian master equation
\begin{align}
\dot{\rho}=\Gamma D[ae^{i\theta\sigma_z}]\rho+\kappa D[a]\rho,\label{eq:3}\tag{3}
\end{align}
where the Lindblad superoperator $D[\bullet]\rho=2\bullet\rho\bullet^\dagger-(\bullet^\dagger\bullet\rho+\rho\bullet^\dagger\bullet)$,  the $\kappa$ denotes the dissipation rate of the cavity mode, and $\Gamma$ denotes the nonreciprocal dispersive coupling rate. The dispersive type of nonreciprocal dissipation operator $D[ae^{i\theta\sigma_z}]\rho$ indicates that the cavity mode exerts a unidirectional influence on the phase and frequency shift of the qubit, with no reverse interaction.

The quantum Langevin equations~\cite{lab25,lab26} for the cavity mode ${a}$ and qubit operator ${\sigma}_-$ can be described by
\begin{align}
&\dot{{a}}=-(\kappa+\Gamma){a}+\sqrt{2\kappa}{a}_{in}(t)-i\sqrt{2\Gamma}e^{-i\sigma_z}{b}_{in}(t),\label{eq:4}\tag{4}\\
&\dot{{\sigma}}_-=[\Gamma(e^{i\theta}-1)]a^\dagger a\sigma_-+2i\sqrt{2\Gamma}\sinh(\frac{i\theta}{2})b_{in}^\dagger(t) \sigma_-a \nonumber\\
&\ \ +2i\sqrt{2\Gamma}\sinh(\frac{-i\theta}{2})a^\dagger \sigma_-b_{in}(t),\label{eq:5}\tag{5}
\end{align}
where ${a}_{in}(t)$ and $b_{in}(t)(t)$ represent quantum input noise operators satisfying the correlation relations:
\begin{align}
&\langle {\eta}_{in}(t){\eta}_{in}(t')\rangle=0,
\langle {\eta}^\dagger_{in}(t){\eta}^\dagger_{in}(t')\rangle=0,\label{eq:6}\tag{6}\\
&\langle {\eta}_{in}(t){\eta}^\dagger_{in}(t')\rangle=\delta(t-t'),
\langle {\eta}^\dagger_{in}(t){\eta}_{in}(t')\rangle=0,\label{eq:7}\tag{7}
\end{align}
in which, the quantum input noise operators $\eta_{in}(t)=\{{a}_{in}(t),\ b_{in}(t)\}$.

\section{estimation of the cavity photon number}
In this section, we aim to employ a qubit to measure the photon number inside the cavity.
By solving the quantum Langevin equation, we can obtain the expectation value
\begin{align}
\langle\sigma_-\rangle=\exp[n\Gamma(e^{i\theta}-1)\frac{1-\exp[-2(\kappa+\Gamma)t]}{2(\kappa+\Gamma)}]\langle\sigma_-\rangle_0,\label{eq:8}\tag{8}
\end{align}
where the initial cavity photon number $n=\langle a^\dagger_0a_0\rangle$ is the parameter to be estimated, and $\langle\sigma_-\rangle_0$ represents the expectation value at the initial moment $t=0$.
When the evolution time reaches the characteristic time $\tau=\frac{1}{\kappa+\Gamma}$, the system is regarded as being in the steady state.
When the initial state of the qubit is $(|0\rangle+|1\rangle)/\sqrt{2}$, the expectation value of the Pauli measurement operator $\sigma_{\varphi}$ at reference phase $\varphi$ is given by
\begin{align}
\langle \sigma_{\varphi}\rangle_s&=
\langle\sigma_-e^{i\varphi}+\sigma_+e^{-i\varphi}\rangle_s\label{eq:9}\tag{9}
\\&=e^\frac{n\Gamma(\cos \theta-1)}{2\kappa+2\Gamma}\cos [\frac{n\Gamma\sin\theta}{2\kappa+2\Gamma}+\varphi],\label{eq:10}\tag{10}
\end{align}
where $\langle\bullet\rangle_s$ denotes the expectation value in the steady state.

The measurement uncertainty can be obtained through the error propagation formula
\begin{align}
\delta^2 n&=\frac{\langle \sigma^2_{\varphi}\rangle_s-\langle \sigma_{\varphi}\rangle_s^2}{|\partial_n\langle \sigma_{\varphi}\rangle_s|^2}\label{eq:11}\tag{11}\\
&=\frac{1-e^\frac{2n\Gamma(\cos \theta-1)}{2\kappa+2\Gamma}\cos ^2 [\frac{n\Gamma\sin(\theta)}{2\kappa+2\Gamma}+\varphi]}{\frac{\Gamma^2}{(2\kappa+2\Gamma)^2}e^{-\frac{n\Gamma(1-\cos(\theta))}{\kappa+\Gamma}}(\cos[\vartheta]-\cos[\vartheta+\theta])^2},\label{eq:12}\tag{12}
\end{align}
where the abbreviation $\partial_n=\frac{\partial}{\partial n}$, and the parameter $\vartheta$ is defined as
\begin{align}
\vartheta=\frac{n\Gamma\sin(\theta)}{2\kappa+2\Gamma}+\varphi.\label{eq:13}\tag{13}
\end{align}

For cavity photon numbers close to zero, i.e., $n\rightarrow0$, we can obtain the analytical result
\begin{align}
\delta^2 n\approx\frac{2n(\Gamma+\kappa)}{\Gamma}\geq2n.\label{eq:14}\tag{14}
\end{align}

For a sufficiently large cavity photon number $(n\gg1)$, the measurement uncertainty is given by
\begin{align}
\delta^2 n\geq{2n e}.\label{eq:15}\tag{15}
\end{align}

With the reciprocal dispersive coupling, we can obtain the measurement uncertainty of parameter $n$ using the above method.
When $n\rightarrow0$ and $\varphi=0$, the measurement uncertainty is given by
\begin{align}
\delta^2 n\approx\frac{4n(\lambda^2+\kappa^2)}{\lambda^2}
\geq4n,\label{eq:16}\tag{16}
\end{align}
where the condition for the final inequality to hold is $\lambda\gg\kappa$.
For $n\gg1$ and $\lambda\gg\kappa$, we find that
\begin{align}
\delta^2 n&\approx\frac{1}{\frac{ \lambda^2}{4\lambda^2+4\kappa^2}e^{\frac{-n \lambda^2}{\lambda^2+\kappa^2}}}\label{eq:17}\tag{17}\\
&\approx{4}{e^{n }}.\label{eq:18}\tag{18}
\end{align}
\section{Calculation of quantum Fisher information}
In this section, we derive the optimal measurement precision by calculating the quantum Fisher information. In the steady state, the density matrix is described by
\begin{align}
\rho=\frac{1}{2}+\langle\sigma_-\rangle|0\rangle\langle1|+\langle\sigma_-\rangle^*|1\rangle\langle0|.\label{eq:19}\tag{19}
\end{align}

The quantum Fisher information can be calculated by the following formula~\cite{lab27,lab28}
\begin{align}
F_Q[\rho]&=\textmd{Tr}[(\partial_n\rho)^2]+\frac{1}{\textmd{Det}[\rho]}\textmd{Tr}[(\rho\partial_n\rho)^2]\label{eq:20}\tag{20}\\
&=\frac{4|\partial_n\langle\sigma_-\rangle|^2+4(\langle\sigma_+\rangle\partial_n\langle\sigma_-\rangle-\langle\sigma_-\rangle\partial_n\langle\sigma_+\rangle)^2}{1-4|\langle\sigma_-\rangle|^2},\label{eq:21}\tag{21}
\end{align}
where $\textmd{Tr}$ denotes the trace operation and $\textmd{Det}$ stands for the determinant calculation.
In the case of the nonreciprocal dispersive coupling, the quantum Fisher information $F_{nr}[\rho]$ is given by
\begin{align}
F_{nr}[\rho]=\frac{2{\Gamma^2(1-\cos \theta)}e^{\Lambda}-{\Gamma^2\sin ^2 \theta}e^{2\Lambda}}{4(\kappa+\Gamma)^2(1-e^\Lambda)},\label{eq:22}\tag{22}
\end{align}
where the parameter $\Lambda=\frac{n\Gamma(\cos \theta-1)}{\kappa+\Gamma}$.

In the case of the reciprocal dispersive coupling, the quantum Fisher information $F_{r}[\rho]$ is given by
\begin{align}
F_r[\rho]=\frac{{\lambda^2}(\lambda^2+\kappa^2)e^{\lambda'}-{\kappa^2\lambda^2}e^{\lambda'}}{4(\lambda^2+\kappa^2)^2(1-e^{\lambda'})},\label{eq:23}\tag{23}
\end{align}
where the parameter $\lambda'=\frac{-n\lambda^2}{\lambda^2+\kappa^2}$.

When $n\rightarrow 0$ or $n\gg1$, the quantum Fisher information can be further simplified as

\[
F_{nr}[\rho]=\left\{
\begin{array}{cc}
\frac{1}{2n}, & n\rightarrow 0; \\
\frac{1}{2ne}, & n\gg1.
\end{array}
\right.\label{eq:24}\tag{24}
\]
\[
F_{r}[\rho]=\left\{
\begin{array}{cc}
\frac{1}{4n}, & n\rightarrow 0; \\
\frac{1}{4e^n}, & n\gg1.
\end{array}
\right.\label{eq:25}\tag{25}
\]

According to the quantum Cram\`{e}r-Rao bound~\cite{lab29,lab30,lab31}, the measurement uncertainties in nonreciprocal ($\delta^2 n_{nr}$) and reciprocal ($\delta^2 n_{r}$) cases can be obtained
\[
\delta^2 n_{nr}\geq \frac{1}{F_{nr}[\rho]}=\left\{
\begin{array}{cc}
{2n}, & n\rightarrow 0; \\
{2ne}, & n\gg1.
\end{array}
\right.\label{eq:26}\tag{26}
\]
\[
\delta^2 n_r\geq \frac{1}{F_{r}[\rho]}=\left\{
\begin{array}{cc}
{4n}, & n\rightarrow 0; \\
{4e^n}, & n\gg1.
\end{array}
\right.\label{eq:27}\tag{27}
\]

These results recover the findings obtained by the measurement operator $\sigma_\varphi$. It shows that the Pauli operator is the optimal measurement operator.
\section{the advantage of the nonreciprocal dispersive coupling}
In this section, we analyze the advantages of nonreciprocal dispersive coupling relative to its reciprocal counterpart under two scenarios: one ignoring time-resource constraints and the other with time resources included.

Without considering time resources, the ratio of the measurement uncertainty from nonreciprocal dispersive coupling to that from its reciprocal counterpart is given by
\[
\delta n_{r}/\delta n_{nr}= \sqrt{F_{nr}[\rho]/F_{r}[\rho]}=\left\{
\begin{array}{cc}
{\sqrt{2}}, & n\rightarrow 0; \\
{\sqrt{2e^{n-1}/n}}, & n\gg1.
\end{array}
\right.\label{eq:28}\tag{28}
\]

\begin{figure}[h]
\includegraphics[scale=0.5]{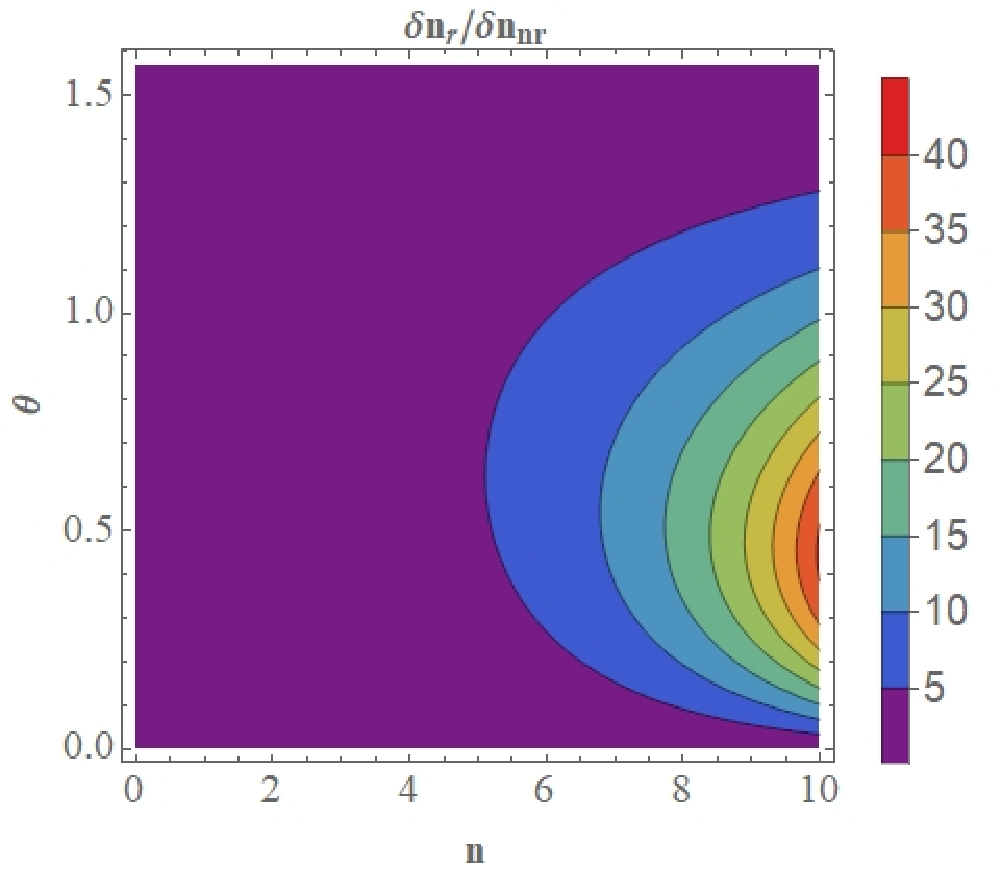}
 \caption{\label{fig.2} Ratio of measurement uncertainty $\delta n_{r}/\delta n_{nr}$ versus cavity photon number $n$ and angle $\theta$ under reciprocal and nonreciprocal dispersive coupling. Here, we consider that $\lambda\gg\kappa$ and $\Gamma\gg\kappa$.}
\end{figure}

This result indicates that when measuring extremely low cavity photon numbers, nonreciprocal dispersive coupling can improves the measurement precision by a factor of $\sqrt{2}$ compared with reciprocal case. For sufficiently large cavity photon numbers, nonreciprocal dispersive coupling achieves an exponential improvement in measurement precision compared with reciprocal case. As shown in Fig.~(\ref{fig.2}), by tuning the angle $\theta$, the measurement uncertainty achieved via nonreciprocal dispersive coupling decreases remarkably as the cavity photon number $n$ increases, compared with that obtained from reciprocal dispersive coupling.

We now discuss how finite time cost impacts the achievable measurement precision. The characteristic times to reach the steady state for reciprocal ($\tau_{r}$) and nonreciprocal ($\tau_{nr}$) dispersive coupling systems are given by
\begin{align}
\tau_r=1/\kappa, \
\tau_{nr}=1/(\kappa+\Gamma).\label{eq:29}\tag{29}
\end{align}
The measurement uncertainties under a given total measurement time $T$ can be expressed as
\begin{align}
\delta n_{rT}\geq\frac{1}{\sqrt{\frac{T}{\tau_r}F_r[\rho]}},\
\delta n_{nrT}\geq\frac{1}{\sqrt{\frac{T}{\tau_{nr}}F_{nr}[\rho]}},\label{eq:30}\tag{30}
\end{align}
where $\delta n_{rT}$ ($\delta n_{nrT}$) denotes the measurement uncertainty with the reciprocal (nonreciprocal) dispersive coupling. The ratio of the measurement uncertainty for the reciprocal dispersive coupling to that for the nonreciprocal counterpart reads
\begin{align}
\frac{\delta n_{rT}}{\delta n_{nrT}}=\sqrt{\frac{F_{nr}[\rho]\tau_r}{F_{r}[\rho]\tau_{nr}}}=\sqrt{\frac{\kappa+\Gamma}{\kappa}}\frac{\delta n_r}{\delta n_{nr}}.\label{eq:31}\tag{31}
\end{align}

This indicates that when time cost is taken into account, nonreciprocal dispersive coupling reaches the equilibrium state faster than its reciprocal counterpart, enabling more measurements to be completed within the given time.
\section{Measurement of driving strength}
In this section, we attempt to measure the single-photon driving strength and explore the role of nonreciprocal dispersive coupling in improving  its measurement precision.

We consider the cavity system driven by a resonant single-photon driving field, and the corresponding Hamiltonian in the rotating frame is given by
\begin{align}
H_d=\varepsilon(a^\dagger+a),\label{eq:32}\tag{32}
\end{align}
where $\varepsilon$ denotes the single-photon driving strength to be measured.
In the case of nonreciprocal dispersive coupling, the evolution equation of the density matrix is described by
\begin{align}
\dot{\rho}=-i[H_d,\rho]+\Gamma D[e^{i\theta/2 \sigma_z}a]\rho+\kappa D[a]\rho.\label{eq:33}\tag{33}
\end{align}
When the initial state of the cavity mode is the vacuum state and the qubit is prepared in the coherent superposition state $(|0\rangle+|1\rangle)/\sqrt{2}$, we can obtain the optimal measurement uncertainty of $\varepsilon$ by employing Pauli operators
 \begin{align}
\delta^2 \varepsilon|_{nr}\approx\frac{\varepsilon^2(e^2-1)}{4}.\label{eq:34}\tag{34}
\end{align}
Under time-resource constraints, the optimal measurement uncertainty of $\varepsilon$ is given by
\begin{align}
\delta^2 \varepsilon|_{nrT}&\approx\frac{(\kappa+\Gamma)^2(e-1)}{4T\Gamma}, \varepsilon\ll1,\label{eq:35}\tag{35}\\
\delta^2 \varepsilon|_{nrT}&=\frac{9(e^{5/3}-1)3^{1/3}\varepsilon^{4/3}}{25(4\Gamma )^{1/3}},  \varepsilon\gg1,\label{eq:36}\tag{36}
\end{align}
where $T$ is the total measurement time.

With the reciprocal dispersive coupling, the optimal measurement uncertainty of $\varepsilon$ is given by
 \begin{align}
\delta^2 \varepsilon|_r&\approx\frac{\varepsilon^2(e^2-1)\kappa^2\lambda^2}{(\kappa^2+\lambda^2)^2}, \varepsilon\ll1,\label{eq:37}\tag{37}\\
\delta^2 \varepsilon|_r&\approx\frac{9(e^{3/2}-1)\varepsilon\lambda}{4*3^{3/2}}, \varepsilon\gg1.\label{eq:38}\tag{38}
\end{align}
Given the fixed total measurement time $T$, the optimal measurement uncertainty of $\varepsilon$ is given by
\begin{align}
\delta^2 \varepsilon|_{rT}&\approx{\kappa(e-1)}/T, \varepsilon\ll1,\label{eq:39}\tag{39}\\
\delta^2 \varepsilon|_{rT}&\approx\frac{9(e^{5/4}-1)\sqrt{\varepsilon\lambda}}{10*(5/2)^{1/4}},  \varepsilon\gg1.\label{eq:40}\tag{40}
\end{align}
By comparing the measurement uncertainties derived from reciprocal and nonreciprocal dispersive coupling in the above formulas, we find that nonreciprocal dispersive coupling exhibits poorer performance, especially at large driving strength.

Next, we will use a new strategy to measure the single-photon driving strength. Firstly, the information of the driving strength is encoded into the cavity photon number. Then, the information is transferred to the qubit via dispersive coupling. Finally, we retrieve the relevant information of driving strength by performing Pauli measurement.

The quantum Langevin equation for the cavity mode ${a}$ is governed by
\begin{align}
\dot{{a}}=-\kappa a-i \varepsilon+\sqrt{2\kappa}{a}_{in}(t).\label{eq:41}\tag{41}
\end{align}

In the steady state, the cavity mode ${a}$ is given by
\begin{align}
{{a}}=\frac{-i\varepsilon}{\kappa}+\int_{t=0}^\infty dt'\sqrt{2\kappa}\exp[-\kappa(t-t')]{a}_{in}(t),\label{eq:42}\tag{42}
\end{align}
Under this condition, the cavity photon number is given by
\begin{align}
n=\langle a^\dagger a\rangle=\frac{\varepsilon^2}{\kappa^2}.\label{eq:43}\tag{43}
\end{align}

Using the new strategy, the measurement uncertainty $\delta \varepsilon|_N$ of the driving strength $\varepsilon$ is derived by  the error propagation formula

\begin{align}
\delta \varepsilon|_N=\frac{\kappa^2}{2\varepsilon}\delta n, \label{eq:44}\tag{44}
\end{align}
where the subscript $N$ represents the new strategy.
Utilizing the measurement uncertainty $\delta n$ in the previous section, we can obtain the measurement uncertainty of the driving strength $\varepsilon$.
For the nonreciprocal dispersive coupling, the corresponding uncertainty $\delta \varepsilon_{nrN}$ achieved by the new strategy is given by
\[
\delta \varepsilon_{nrN} =
\left\{
\begin{array}{ll}
{\sqrt{2}\kappa}/{2}, & \varepsilon\ll\kappa; \\
{\sqrt{2e}\kappa}/{2}, & \varepsilon\gg\kappa.
\end{array}
\right.\label{eq:45}\tag{45}
\]
The total characteristic time $\tau$ reads
\begin{align}
\tau=\tau_1+\tau_2=\frac{1}{\kappa}+\frac{1}{\kappa+\Gamma},\label{eq:46}\tag{46}
\end{align}
where $\tau_1$ represents the characteristic time for encoding the driving strength onto the cavity state; $\tau_2$ represents the characteristic time for the qubit to reach the steady state.
Given the total measurement time $T$, the number of repeated measurements is $T/\tau$. After a given measurement time $T$, the measurement uncertainty $\delta^2 \varepsilon_{nrNT}$	for nonreciprocal dispersive coupling under the new strategy is thus given by
\[
\delta^2 \varepsilon_{nrNT} =
\left\{
\begin{array}{ll}
\frac{\kappa(2\kappa+\Gamma)}{2(\kappa+\Gamma)T}, & \varepsilon\ll\kappa; \\
\frac{e\kappa(2\kappa+\Gamma)}{2(\kappa+\Gamma)T}, & \varepsilon\gg\kappa.
\end{array}
\right.\label{eq:47}\tag{47}
\]
With the new strategy, the measurement uncertainty $\delta \varepsilon_{rN}$ for reciprocal dispersive coupling is given by
\[
\delta \varepsilon_{rN} =
\left\{
\begin{array}{ll}
\kappa, & \varepsilon\ll\kappa; \\
e^{\varepsilon^2/(2\kappa^2)}\kappa^2/\varepsilon, & \varepsilon\gg\kappa.
\end{array}
\right.\label{eq:48}\tag{48}
\]

Given the measurement time $T$, the number of repeated measurements is $T/\tau$, where the total characteristic time is
\begin{align}
\tau=\tau_1+\tau_2=1/\kappa+1/\kappa=2/\kappa.\label{eq:49}\tag{49}
\end{align}
Then, taking the cost of the measurement time $T$ into account, we can achieve the measurement uncertainty $\delta^2 \varepsilon_{rNT}$ obtained by the reciprocal dispersive coupling under the new strategy
\[
\delta^2 \varepsilon_{rNT} =
\left\{
\begin{array}{ll}
2\kappa/T, & \varepsilon\ll\kappa; \\
2e^{\varepsilon^2/(\kappa^2)}\kappa/(T\varepsilon), & \varepsilon\gg\kappa.
\end{array}
\right.\label{eq:50}\tag{50}
\]

By comparing $\delta \varepsilon_{rN}$ in Eq.~(\ref{eq:48}) with $\delta \varepsilon_{nrN}$ in Eq.~(\ref{eq:45}) as well as $\delta^2 \varepsilon_{rNT}$ in  Eq.~(\ref{eq:50}) with $\delta^2 \varepsilon_{nrNT}$ in  Eq.~(\ref{eq:47}), we find clearly that nonreciprocal dispersive coupling always achieves higher measurement precision than its reciprocal counterpart under this new strategy, regardless of the time cost.

Next, we compare the optimal measurement uncertainty achieved by the new strategy with that obtained via the high-performance reciprocal dispersive coupling in the previous direct measurement scheme.
We find that
\[
\delta^2 \varepsilon_{rT}/\delta^2 \varepsilon_{nrNT} =
\left\{
\begin{array}{ll}
(e-1)\approx1.7, & \varepsilon\ll\kappa; \\
\frac{9(e^{5/4}-1)\sqrt{\varepsilon\lambda}}{10*(5/2)^{1/4}e\kappa}\approx0.7\sqrt{\varepsilon\lambda}/\kappa, & \varepsilon\gg\kappa.
\end{array}
\right.\label{eq:51}\tag{51}
\]
This result shows that under this new strategy, nonreciprocal dispersive coupling maintains higher measurement precision than the reciprocal counterpart. Moreover, its advantages become more pronounced at larger driving strengths.
\section{Conclusion}
We have explored the role of nonreciprocal dispersive coupling in quantum sensing. A typical nonreciprocal dispersive coupling can be constructed by adiabatically eliminating an intermediate bosonic mode with strong dissipation. Most existing works employ optical cavities to extract information from qubits. In contrast, we adopt the opposite approach and employ qubits to read out information encoded in cavity fields. The Pauli measurement operators acting on qubits can be implemented via Ramsey measurements~\cite{lab16}. By comparison with the measurement precision obtained by the quantum Fisher information, we demonstrate that the Pauli operator is the optimal measurement.
When measuring the cavity photon number, nonreciprocal dispersive coupling delivers higher precision than its reciprocal counterpart, and the ratio of their precision grows exponentially with the cavity photon number. After accounting for measurement time, the precision of nonreciprocal dispersive coupling can be further improved. However, when qubits are used to directly measure the single-photon driving strength, nonreciprocal dispersive coupling shows no obvious advantages.

We adopt a novel measurement strategy: first convert the information of single-photon driving strength into cavity photon numbers, and then transmit such information to qubits via nonreciprocal dispersive coupling. Under this novel strategy, nonreciprocal dispersive coupling achieves higher precision than  the reciprocal counterpart in both measurement schemes (the direct measurement and the proposed strategy).

Our work paves the way for developing high-precision quantum sensing devices based on nonreciprocal dispersive coupling. Extending this coupling from qubit-cavity field systems to architectures with high-dimensional qudits~\cite{lab32,lab33,lab34} and general quantum fields represents a promising direction for future research.

\section*{Acknowledgements}
This research was supported by the National Natural Science Foundation of China (Grant No. 12365001), and the Bagui Youth Top Talent Training Program.
\newpage
\section*{Appendix A: Construction of non-reciprocal dispersive coupling}
The quantum Langevin equations of the cavity mode ${a}$, bosonic mode $b$, and qubit mode ${\sigma_-}$ can be described by
\begin{align}
&\dot{{a}}=(-i\Delta-\kappa)a-igb +\sqrt{2\kappa}{a}_{in}(t),\label{eq:A1}\tag{A1}\\
&\dot{{b}}=-\kappa_bb-iga-i\chi_0\sigma_zb+\sqrt{2\kappa_b}{b}_{in}(t),\label{eq:A2}\tag{A2}\\
&\dot{{\sigma_-}}=-i\chi b^\dagger \sigma_- b.\label{eq:A3}\tag{A3}
\end{align}
When $\kappa_b\gg\kappa$, the bosonic mode $b$ can be eliminated adiabatically.
\begin{align}
\dot{{b}}=0\Rightarrow
b=\frac{-iga +\sqrt{2\kappa_b}{b}_{in}(t)}{\kappa_b+i\chi \sigma_z}.\label{eq:A4}\tag{A4}
\end{align}
Substituting the above result into Eq.~(\ref{eq:A1}), we can obtain the evolution equation of the cavity mode $a$
\begin{align}
\dot{a}=&-[i (\Delta-\lambda)\sigma_z+(\kappa+\Gamma)]a\nonumber\\
&+\sqrt{2\kappa}a_{in}(t)-i\sqrt{2\Gamma}e^{-i\frac{\theta}{2}\sigma_z}b_{in}(t),\label{eq:A5}\tag{A5}
\end{align}
in which,
\begin{align}
&\Gamma=\frac{g^2\kappa_b}{\kappa_b^2+\chi ^2},\ \lambda=\frac{g^2\chi }{\kappa_b^2+\chi ^2},\label{eq:A6}\tag{A6}\\
&\cos(\theta/2)=\frac{\kappa_b}{\sqrt{\kappa_b^2+\chi ^2}},\ \sin(\theta/2)=\frac{\chi }{\sqrt{\kappa_b^2+\chi ^2}}.\label{eq:A7}\tag{A7}
\end{align}
Substituting Eq.~(\ref{eq:A4}) into Eq.~(\ref{eq:A3}), we can obtain the evolution equation of the qubit $\sigma_-$
\begin{align}
&\dot{{\sigma_-}}=[i\lambda+\Gamma(e^{i\theta}-1)]a^\dagger a\sigma_--i2\sqrt{2\Gamma}\sinh(\frac{i\theta}{2})b_{in}^\dagger(t) \sigma_-a \nonumber\\
&+i2\sqrt{2\Gamma}\sinh(\frac{-i\theta}{2})a^\dagger \sigma_-b_{in}(t)-i\sin\theta e^{-i\theta}b^\dagger_{in}(t)\sigma_-b_{in}(t).\label{eq:A8}\tag{A8}
\end{align}

Based on the above quantum Langevin equations, the corresponding evolution equation of the density matrix is
\begin{align}
\dot{\rho}=-i[H_e,\rho]+\Gamma D[e^{i\theta/2 \sigma_z}a]\rho+\kappa D[a]\rho,\label{eq:A9}\tag{A9}
\end{align}
where the effective Hamiltonian is
\begin{align}
H_e=\Delta a^\dagger a-\lambda a^\dagger a \sigma_z.\label{eq:A10}\tag{A10}
\end{align}
To achieve fully nonreciprocal dispersive coupling, coherent dispersive coupling between the cavity and qubit is required
\begin{align}
H_c=\lambda_c a^\dagger a\sigma_z.\label{eq:A11}\tag{A11}
\end{align}
When $\lambda_c=\lambda$, the nonreciprocal dispersive coupling between the cavity mode and qubit is achieved. The evolution of the density matrix is described by
\begin{align}
\dot{\rho}=-i[\Delta a^\dagger a,\rho]+\Gamma D[e^{i\theta/2 \sigma_z}a]\rho+\kappa D[a]\rho. \label{eq:A12}\tag{A12}
\end{align}
Without loss of generality, we consider the resonant case in the main text, i.e., $\Delta=0$.
\section*{Appendix B: measurement uncertainty with nonreciprocal dispersive coupling}
For the nonreciprocal dispersive coupling, the quantum Langevin equations for the cavity mode ${a}$ and qubit ${\sigma}_-$ can be described by
\begin{align}
\dot{{a}}=&-(\kappa+\Gamma){a}+\sqrt{2\kappa}{a}_{in}(t)-i\sqrt{2\Gamma}e^{-i\sigma_z}{b}_{in}(t),\label{eq:B1}\tag{B1}\\
\dot{{\sigma}}_-=&[\Gamma(e^{i\theta}-1)]a^\dagger a\sigma_-+2i\sqrt{2\Gamma}\sinh(\frac{i\theta}{2})b_{in}^\dagger(t) \sigma_-a \nonumber\\
&+2i\sqrt{2\Gamma}\sinh(\frac{-i\theta}{2})a^\dagger \sigma_-b_{in}(t).\label{eq:B2}\tag{B2}
\end{align}
Because the operator $\sigma_z$ remains unchanged during the evolution process, the analytical result can be achieved
\begin{align}
 &a=\exp[-(\kappa+\Gamma)t]a_0+\int_{0}^tdt'\exp[-(\kappa+\Gamma)(t-t')]c_{in}(t'),\label{eq:B3}\tag{B3}\\
&\langle\sigma_-\rangle=\exp[\langle a^\dagger_0a_0\rangle\Gamma(e^{i\theta}-1)\frac{1-\exp[-2(\kappa+\Gamma)t]}{2(\kappa+\Gamma)}]\langle\sigma_-\rangle_0,\label{eq:B4}\tag{B4}
\end{align}
where the noise operator $c_{in}(t')=\sqrt{2\kappa}a_{in}(t')+\sqrt{2\Gamma}b_{in}(t')$, and $a_0$ represents the operator $a$ at the initial moment $t=0$.

When the system reach the steady state, the expectation value of the Pauli operator $\sigma_\varphi$ is given by
\begin{align}
\langle \sigma_{\varphi}\rangle=&
\langle\sigma_-e^{i\varphi}+\sigma_+e^{-i\varphi}\rangle\label{eq:B5}\tag{B5}\\
=&e^\frac{n\Gamma(\cos \theta-1)}{2\kappa+2\Gamma}\cos [\frac{n\Gamma\sin(\theta)}{2\kappa+2\Gamma}+\varphi].\label{eq:B6}\tag{B6}
\end{align}
By utilizing the above result and $\langle \sigma_{\varphi}^2\rangle=1$, the measurement uncertainty of the cavity photon number $n$ can be derived by the error propagation formula
\begin{align}
\delta^2 n&=\frac{\langle \sigma^2_{\varphi}\rangle-\langle \sigma_{\varphi}\rangle^2}{|\partial_n\langle \sigma_{\varphi}\rangle|^2}\label{eq:B7}\tag{B7}\\
&=\frac{1-e^\frac{2n\Gamma(\cos \theta-1)}{2\kappa+2\Gamma}\cos ^2 [\vartheta]}{\frac{\Gamma^2}{(2\kappa+2\Gamma)^2}e^{-\frac{n\Gamma(1-\cos(\theta))}{\kappa+\Gamma}}(\cos[\vartheta]-\cos[\vartheta+\theta])^2},\label{eq:B8}\tag{B8}
\end{align}
where the parameter $\vartheta$ is defined as
\begin{align}
\vartheta=\frac{n\Gamma\sin(\theta)}{2\kappa+2\Gamma}+\varphi.\label{eq:B9}\tag{B9}
\end{align}

When $n\rightarrow0$, we choose the phases $\theta=\pi$ and $\varphi=0$ to obtain the minimum measurement uncertainty.
\begin{align}
\delta^2 n=&\frac{1-e^\frac{-2n\Gamma}{\kappa+\Gamma}}{\frac{\Gamma^2}{(\kappa+\Gamma)^2}e^{-\frac{2n\Gamma}{\kappa+\Gamma}}}\approx\frac{1-(1-\frac{-2n\Gamma}{\kappa+\Gamma})}{\frac{\Gamma^2}{(\kappa+\Gamma)^2}}\label{eq:B10}\tag{B10}\\
\approx&\frac{2n(\Gamma+\kappa)}{\Gamma}\geq2n,\label{eq:B11}\tag{B11}
\end{align}
where the condition for obtaining the final minimum value is $\kappa\ll\Gamma$.

For $n\gg1$,  the minimum measurement uncertainty of $n$ is derived
\begin{align}
\delta^2 n=&\frac{1}{e^{-n(1-\cos \theta) }\sin ^2 \frac{\theta}{2}\sin ^2 (\vartheta+\frac{\theta}{2})}\label{eq:B12}\tag{B12}\\
\geq&\frac{1}{e^{-n(1-\cos \theta) }\sin ^2 \frac{\theta}{2}}\label{eq:B13}\tag{B13}\\
\geq&{2n e}.\label{eq:B14}\tag{B14}
\end{align}
where we have chosen $\vartheta+\frac{\theta}{2}=\pi/2$.
The condition for the last equation to hold is
\begin{align}
\sin ^2 \frac{\theta}{2}=\frac{1}{2n}.\label{eq:B15}\tag{B15}
\end{align}
When $n\rightarrow\infty$, we can derive that $\theta\rightarrow0$ and $\varphi=\pi/2$.

\section*{Appendix C: measurement uncertainty with reciprocal dispersive coupling}
For the reciprocal dispersive coupling, the evolution equation of the density matrix is given by
\begin{align}
\dot{\rho}=-i[\lambda a^\dagger a,\rho]+\kappa D[a]\rho.\label{eq:C1}\tag{C1}
\end{align}

The quantum Langevin equation of the cavity mode ${a}$ and qubit ${\sigma}_-$ can be described by
\begin{align}
&\dot{{a}}=-(\kappa+i\lambda \sigma_z){a}+\sqrt{2\kappa}{a}_{in}(t),\label{eq:C2}\tag{C2}\\
&\dot{{\sigma}}_-=-i\lambda a^\dagger \sigma_-a,\label{eq:C3}\tag{C3}
\end{align}

The analytical solutions are derived
\begin{align}
 a=&\exp[-i (\lambda\sigma_z-i\kappa)t]a_0\nonumber\\
 &+\int_{0}^tdt'\exp[-i (\lambda\sigma_z-i\kappa)(t-t')]\sqrt{2\kappa}a_{in}(t')\label{eq:C4}\tag{C4}\\
\langle\sigma_-\rangle=&\exp[i \lambda n\frac{\exp[2(i\lambda-\kappa)t]-1}{2i\lambda-2\kappa}]\langle\sigma_-\rangle_0.\label{eq:C5}\tag{C5}
\end{align}
The expectation value of the Pauli operator is given by
\begin{align}
\langle \sigma_{\varphi}\rangle&=\langle\sigma_-e^{i\varphi}+\sigma_+e^{-i\varphi}\rangle\label{eq:C6}\tag{C6}\\
&=\frac{1}{2}(\exp[i \lambda n \frac{\exp[2(i\lambda-\kappa)t]-1}{2i\lambda-2\kappa}+i\varphi]+h.c.).\label{eq:C7}\tag{C7}
\end{align}

According to the error propagation formula, the measurement uncertainty of the cavity photon number is
\begin{align}
\delta^2 n&=\frac{\langle \sigma^2_{\varphi}\rangle-\langle \sigma_{\varphi}\rangle^2}{|\partial_n\langle \sigma_{\varphi}\rangle|^2}\label{eq:C8}\tag{C8}\\
&=\frac{1-e^{\frac{-n \lambda^2}{\lambda^2+\kappa^2}}\cos^2[\frac{n \lambda^2}{2\lambda^2+2\kappa^2}+\varphi]}{\frac{ \lambda^2}{4\lambda^2+4\kappa^2}e^{\frac{-n \lambda^2}{\lambda^2+\kappa^2}}\sin^2(\varphi+\frac{\kappa\lambda n}{2\lambda^2+2\kappa^2}+\phi)},\label{eq:C9}\tag{C9}
\end{align}
where the parameter $\phi$ satisfies
\begin{align}
\sin \phi=\frac{\lambda}{\sqrt{\lambda^2+\kappa^2}}.\label{eq:C10}\tag{C10}
\end{align}

When $n\rightarrow0$, we obtain the measurement uncertainty with the Pauli operator $\sigma_x$
\begin{align}
\delta^2 n&=\frac{1-e^{\frac{-n \lambda^2}{\lambda^2+\kappa^2}}}{\frac{ \lambda^2}{4\lambda^2+4\kappa^2}e^{\frac{-n \lambda^2}{\lambda^2+\kappa^2}}\sin^2(\phi)}\label{eq:C11}\tag{C11}\\
&\approx\frac{4n(\lambda^2+\kappa^2)}{\lambda^2}
\geq4n,\label{eq:C12}\tag{C12}
\end{align}
where the condition for the final inequality to hold is $\lambda\gg\kappa$.

For $n\gg1$ and $\lambda\gg\kappa$, we can obtain that
\begin{align}
\delta^2 n&\approx\frac{1}{\frac{ \lambda^2}{4\lambda^2+4\kappa^2}e^{\frac{-n \lambda^2}{\lambda^2+\kappa^2}}}\label{eq:C13}\tag{C13}\\
&\approx{4}{e^{n }}.\label{eq:C14}\tag{C14}
\end{align}

\section*{Appendix D: Measurement of driving strength}
For the nonreciprocal dispersive coupling, the evolution equation in the rotating frame reads
\begin{align}
\dot{\rho}=-i[H_d,\rho]+\Gamma D[e^{i\theta/2 \sigma_z}a]\rho+\kappa D[a]\rho,\label{eq:D1}\tag{D1}
\end{align}
where the Hamiltonian of the single-photon drive is
\begin{align}
H_d=\varepsilon(a^\dagger+a).\label{eq:D2}\tag{D2}
\end{align}
By solving the corresponding quantum Langevin equation, the analytical results are obtained
\begin{align}
 a=&\alpha(t)+\exp[-(\kappa+\Gamma)t]a_0\nonumber\\
 &+\int_{0}^tdt'\exp[-(\kappa+\Gamma)(t-t')]c_{in}(t'),\label{eq:D3}\tag{D3}\\
\textmd{ in which},\nonumber\\
 \alpha(t)&=\frac{-i\varepsilon(1-\exp[-(\kappa+\Gamma)t])}{\kappa+\Gamma},\label{eq:D4}\tag{D4}\\
 c_{in}(t')&=\sqrt{2\kappa}a_{in}(t')+\sqrt{2\Gamma}b_{in}(t').\label{eq:D5}\tag{D5}
\end{align}
The evolution equation with respect to the expectation value of $\sigma_-$ is given by
\begin{align}
\langle\dot{\sigma}_-\rangle&=\Gamma(e^{i\theta}-1)\langle (a^\dagger_0+ \alpha^*(t))(a_0+ \alpha(t))\rangle\langle\sigma_-\rangle.\label{eq:D6}\tag{D6}
\end{align}
When the cavity mode is initially in the vacuum state, solving the above equation yields the expectation value of $\sigma_-$
\begin{align}
\langle{\sigma}_-\rangle=&\exp[\frac{\Gamma(e^{i\theta}-1)\varepsilon^2}{2(\kappa+\Gamma)^3}(2\kappa t+2\Gamma t-3-e^{-2(\kappa+\Gamma)t}\nonumber\\
&+4e^{-(\kappa+\Gamma)t})]\langle{\sigma}_-\rangle_0.\label{eq:D7}\tag{D7}
\end{align}

When the initial state of the qubit is in the coherent superposition state $(|0\rangle+|1\rangle)/\sqrt{2}$, the expectation value of the Pauli operator $\sigma_\varphi$ is
\begin{align}
\langle{\sigma}_\varphi\rangle=&\frac{1}{2}\exp\{\frac{\Gamma(e^{i\theta}-1)\varepsilon^2}{2(\kappa+\Gamma)^3}[2\kappa t+2\Gamma t-3-e^{-2(\kappa+\Gamma)t}\nonumber\\
&+4e^{-(\kappa+\Gamma)t}]+i\varphi\}+h.c.\label{eq:D8}\tag{D8}
\end{align}

According to the error propagation formula, the measurement uncertainty of the driving strength is obtained for $\theta=\pi$
\begin{align}
\delta^2 \varepsilon&=\frac{\langle \sigma^2_{\varphi}\rangle-\langle \sigma_{\varphi}\rangle^2}{|\partial_\varepsilon\langle \sigma_{\varphi}\rangle|^2}
=\frac{1-\langle \sigma_{\varphi}\rangle^2}{\mu^2|\langle \sigma_{\varphi}\rangle|^2}\label{eq:D9}\tag{D9}\\
&=\frac{1-\cos ^2\varphi e^{-2\mu\varepsilon^2}}{(2\mu \varepsilon)^2\cos ^2\varphi e^{-2\mu\varepsilon^2}}
\geq\frac{1- e^{-2\mu\varepsilon^2}}{(2\mu \varepsilon)^2 e^{-2\mu\varepsilon^2}}\label{eq:D10}\tag{D10}\\
&\textmd{in which},\nonumber\\
\mu&=\frac{\Gamma}{(\kappa+\Gamma)^3}[2\kappa t+2\Gamma t-3-e^{-2(\kappa+\Gamma)t}+4e^{-(\kappa+\Gamma)t}].\label{eq:D11}\tag{D11}
\end{align}
When $\mu=1/\varepsilon^2$, we can obtain the optimal measurement uncertainty
 \begin{align}
\delta^2 \varepsilon\approx\frac{\varepsilon^2(e^2-1)}{4}.\label{eq:D12}\tag{D12}
\end{align}
The optimal measurement time is given by the solution of the following equation
 \begin{align}
\frac{\Gamma}{(\kappa+\Gamma)^3}[2\kappa t+2\Gamma t-3-e^{-2(\kappa+\Gamma)t}+4e^{-(\kappa+\Gamma)t}]=\frac{1}{\varepsilon^2}.\label{eq:D13}\tag{D13}
\end{align}
When $\varepsilon\ll 1 $, the optimal single measurement time is given by
 \begin{align}
t_o\approx\frac{(\kappa+\Gamma)^2}{2\Gamma \varepsilon^2}.\label{eq:D14}\tag{D14}
\end{align}
When $\varepsilon\gg 1 $, the optimal measurement time $t$ will be close to 0. Then, we can derive that
 \begin{align}
\frac{\Gamma}{(\kappa+\Gamma)^3}[2\kappa t+2\Gamma t-3-e^{-2(\kappa+\Gamma)t}+4e^{-(\kappa+\Gamma)t}]
\approx\frac{2}{3}\Gamma t^3.\label{eq:D15}\tag{D15}
\end{align}
The optimal measurement time $t$ is given by
 \begin{align}
t_o=(\frac{3}{2\Gamma \varepsilon^2})^{1/3}.\label{eq:D16}\tag{D16}
\end{align}

Taking into account the cost of the measurement time, the measurement uncertainty is expressed as
\begin{align}
\delta^2 \varepsilon|_t\geq\frac{t(1- e^{-2\mu\varepsilon^2)}}{T(2\mu \varepsilon)^2 e^{-2\mu\varepsilon^2}},\label{eq:D17}\tag{D17}
\end{align}
where $T$ is the total measurement time.
When $\varepsilon\ll 1 $, the optimal measurement time is given by
 \begin{align}
t_o=\frac{(\kappa+\Gamma)^2}{4\Gamma \varepsilon^2}.\label{eq:D18}\tag{D18}
\end{align}
The corresponding measurement uncertainty of $\varepsilon$ is
 \begin{align}
\delta^2 \varepsilon|_T\approx\frac{(\kappa+\Gamma)^2(e-1)}{4T\Gamma}.\label{eq:D19}\tag{D19}
\end{align}

When $\varepsilon\gg 1 $, the measurement uncertainty with the fixed total measurement time $T$ is given by
 \begin{align}
\delta^2 \varepsilon\approx\frac{1- e^{-4 \Gamma t^3\varepsilon^2/3}}{(4 \Gamma\varepsilon^2/3)^2t^5 e^{-4 t^3\varepsilon^2 \Gamma/3T}}.\label{eq:D20}\tag{D20}
\end{align}
By differentiating time $t$, we can derive the optimal measurement time
 \begin{align}
t_o=(\frac{5}{4\Gamma \varepsilon^2})^{1/3}.\label{eq:D21}\tag{D21}
\end{align}
At this point, the corresponding optimal measurement uncertainty reads
 \begin{align}
\delta^2 \varepsilon|_m=\frac{9(e^{5/3}-1)3^{1/3}\varepsilon^{4/3}}{25(4\Gamma )^{1/3}}.\label{eq:D22}\tag{D22}
\end{align}

\section*{Appendix E: Measurement of driving strength with reciprocal dispersive coupling}
Using similar derivations as presented above, we can obtain the solution in the case of reciprocal dispersive coupling
\begin{align}
 a=&\alpha(t)+\exp[i (\lambda\sigma_z+i\kappa)t]a_0\nonumber\\
 &+\int_{0}^tdt'\exp[i (\lambda\sigma_z+i\kappa)(t-t')]\sqrt{2\kappa}a_{in}(t')\label{eq:E1}\tag{E1}\\
\textmd{ in which}, \nonumber\\
\alpha(t)&=\frac{i \varepsilon}{i\lambda \sigma_z-\kappa}(1-e^{(i\lambda\sigma_z-\kappa)t})\label{eq:E2}\tag{E2}
\end{align}

When the initial state of the cavity mode is the vacuum state, the evolution equation with respect to the expected value of $\sigma_-$ is given by
\begin{align}
\langle\dot{\sigma}_-\rangle=&i\lambda\langle a^\dagger{\sigma}_-a\rangle\label{eq:E3}\tag{E3}\\
=& \frac{i\lambda\varepsilon^2}{(i\lambda-\kappa)^2}(1-e^{(i\lambda-\kappa)t})^2\langle {\sigma}_-\rangle\label{eq:E4}\tag{E4}
\end{align}
The solution of the above equation is
\begin{align}
\langle{\sigma}_-\rangle=&\exp[\frac{i\lambda\varepsilon^2}{2(\kappa-i\lambda)^3}[2\kappa t-2i\lambda t-3-e^{-2(\kappa-i\lambda)t}\nonumber\\
&+4e^{-(\kappa-i\lambda)t}]]\langle{\sigma}_-\rangle_0.\label{eq:E5}\tag{E5}
\end{align}

According to the error propagation formula,  the measurement uncertainty is given by
 \begin{align}
\delta^2 \varepsilon=\frac{1-\langle \sigma_{\varphi}\rangle^2}{|\nu \langle{\sigma}_-\rangle e^{i\varphi}+\nu \langle{\sigma}_-\rangle^*e^{-i\varphi}|^2},\label{eq:E6}\tag{E6}
\end{align}
where the parameter $\nu $ is
\begin{align}
\nu=&\frac{i\lambda\varepsilon}{2(\kappa-i\lambda)^3}\nonumber\\
&\times[2\kappa t-2i\lambda t-3-e^{-2(\kappa-i\lambda)t}+4e^{-(\kappa-i\lambda)t}].\label{eq:E7}\tag{E7}
\end{align}

When $\varepsilon\ll1$, the above equation can be simplified into the form
 \begin{align}
\delta^2 \varepsilon=\frac{1-e^{-\frac{4\lambda^2 \kappa \varepsilon^2 t}{(\kappa^2+\lambda^2)^2}}\cos ^2[\varphi+{\frac{\lambda(\kappa^2-\lambda^2)\varepsilon^2 t}{(\kappa^2+\lambda^2)^2}]}}{|e^{-\frac{2\lambda^2\kappa\varepsilon^2 t}{(\kappa^2+\lambda^2)^2}}\frac{2t\varepsilon\lambda}{(\kappa^2+\lambda^2)}\sin[\varphi+\frac{\lambda(\kappa^2-\lambda^2)\varepsilon^2 t}{(\kappa^2+\lambda^2)^2}+\phi]|^2},\label{eq:E8}\tag{E8}
\end{align}
where the phase $\phi$ satisfies
 \begin{align}
\sin \phi=\frac{-2\kappa\lambda}{\kappa^2+\lambda^2},\
\cos \phi=\frac{\lambda^2-\kappa^2}{\lambda^2+\kappa^2}.\label{eq:E9}\tag{E9}
\end{align}
By taking the derivative with respect to time $t$, we can obtain the optimal measurement time and the corresponding measurement uncertainty
\begin{align}
&t_o=\frac{(\kappa^2+\lambda^2)^2}{2\lambda^2\kappa\varepsilon^2 },\label{eq:E10}\tag{E10}\\
&\delta^2 \varepsilon|_m\approx\frac{\varepsilon^2(e^2-1)\kappa^2\lambda^2}{(\kappa^2+\lambda^2)^2}.\label{eq:E11}\tag{E11}
\end{align}

Including the cost of the measurement time, the optimal measurement time and the corresponding measurement uncertainty are given by
 \begin{align}
&t_o=\frac{(\kappa^2+\lambda^2)^2}{4\lambda^2\kappa\varepsilon^2 },\label{eq:E12}\tag{E12}\\
&\delta^2 \varepsilon|_T\approx{\kappa(e-1)}/T.\label{eq:E13}\tag{E13}
\end{align}

When $\varepsilon\gg1$, the expectation value of the Pauli measurement operator is
\begin{align}
\langle \sigma_\varphi\rangle=e^{-\lambda^2\varepsilon^2t^4/4}\cos(\lambda\varepsilon^2t^3/3+\varphi).\label{eq:E14}\tag{E14}
\end{align}
According to the error propagation formula,  the measurement uncertainty is given by
 \begin{align}
\delta^2 \varepsilon\approx\frac{1-e^{-\lambda^2\varepsilon^2t^4/2}}{\frac{4}{9}e^{-\lambda^2\varepsilon^2t^4/2}t^6\varepsilon^2\lambda^2}.\label{eq:E15}\tag{E15}
\end{align}
Following a similar derivation as above,  the optimal measurement time and the corresponding measurement uncertainty are given by
 \begin{align}
&t_o=\frac{3^{1/4}}{(\varepsilon\lambda)^{1/2}}.\label{eq:E16}\tag{E16}\\
&\delta^2 \varepsilon_m\approx\frac{9(e^{3/2}-1)\varepsilon\lambda}{4 \ast3^{3/2}}.\label{eq:E17}\tag{E17}
\end{align}

Including the cost of the measurement time, the measurement uncertainty is obtained
\begin{align}
\delta^2 \varepsilon\approx\frac{1-e^{-\lambda^2\varepsilon^2t^4/2}}{\frac{4}{9}Te^{-\lambda^2\varepsilon^2t^4/2}t^5\varepsilon^2\lambda^2}.\label{eq:E18}\tag{E18}
\end{align}
The optimal measurement time and the corresponding measurement uncertainty are given by
 \begin{align}
t_o=\frac{(5/2)^{1/4}}{(\varepsilon\lambda)^{1/2}},\label{eq:E19}\tag{E19}
\end{align}
\begin{align}
\delta^2 \varepsilon_m\approx\frac{9(e^{5/4}-1)\sqrt{\varepsilon\lambda}}{10\ast(5/2)^{1/4}}.\label{eq:E20}\tag{E20}
\end{align}

\end{document}